# Explainable Threat Attribution for IoT Networks Using Conditional SHAP and Flow Behavior Modelling

Samuel Ozechi[1]   Jennifer Okonkwoabutu[2]

[1]University of East London, United Kingdom
[2]ECE Ecole D'Ingenieur, France

**Abstract**

As the Internet of Things (IoT) continues to expand across critical infrastructure, smart environments, and consumer devices, securing them against cyber threats has become increasingly vital. Traditional intrusion detection models often treat IoT threats as binary classification problems or rely on opaque models, thereby limiting trust. This work studies multiclass threat attribution in IoT environments using the CICIoT2023 dataset, grouping over 30 attack variants into 8 semantically meaningful classes. We utilize a combination of a gradient boosting model and SHAP (SHapley Additive exPlanations) to deliver both global and class-specific explanations, enabling detailed insight into the features driving each attack classification. The results show that the model distinguishes distinct behavioral signatures of the attacks using flow timing, packet size uniformity, TCP flag dynamics, and statistical variance. Additional analysis that exposes both feature attribution and the decision trajectory per class further validates these observed patterns. Our findings contribute to the development of more accurate and explainable intrusion detection systems, bridging the gap between high-performance machine learning and the need for trust and accountability in AI-driven cybersecurity for IoT environments.

## 1 Introduction

The rapid proliferation of devices on the Internet of Things (IoT) has led to a paradigm shift in the design of modern digital ecosystems. From smart homes and industrial sensors to connected healthcare and critical infrastructure, IoT networks now play a vital role in both consumer and enterprise contexts. However, this widespread connectivity introduces a vastly expanded attack surface, making IoT networks increasingly vulnerable to a variety of cyber threats such as distributed denial of service (DDoS), spoofing, scanning, brute-force intrusions, and malware propagation [1, 2, 3]. The heterogeneous and resource-constrained nature of IoT devices further exacerbates the challenge, as traditional security mechanisms are often too rigid or computationally intensive to implement effectively[2]. Among the most urgent needs in this domain is the ability to perform robust and fine-grained intrusion detection. Many existing systems focus on binary classification (benign vs. malicious), but real-world IoT attacks are diverse and multifaceted[4]. Multiclass intrusion detection, where each class corresponds to a specific type of threat, is much more realistic and informative for forensic analysis and automated defense. However, as the number of attack types grows, so does the complexity of both detection and interpretation. Models trained on large-scale data can achieve high accuracy, but their decision-making processes are often opaque, limiting their adoption in high-stakes environments. This has driven a growing demand for explainable artificial intelligence (XAI) in cybersecurity. In particular, stakeholders need not only accurate predictions but also transparent and actionable explanations of why a flow was classified as, for example, a brute-force attack versus a DDoS event [5]. Traditional feature importance metrics are often too coarse or generic, while black-box deep learning models lack explainability altogether. A comprehensive survey of deep learning applications in cybersecurity reveals [6].

Although recent research has explored XAI techniques for cybersecurity, using technologies such as SHAP (SHapley Additive exPlanations) [7, 8] and LIME [9], these techniques are often applied in binary contexts or fail to capture the per-class decision logic essential for threat attribution. To address these limitations, this work presents a novel explainable threat attribution framework for IoT networks, combining flow behavior analysis and modeling with class-aware SHAP analysis. The CICIoT2023[10], one of the most comprehensive multiclass IoT attack datasets available, was used, and a gradient-boosting model was trained to detect and classify various threat categories. Rather than focusing solely on predictive performance, we focus on understanding how the model makes its decisions, both globally across the dataset and locally across classes. Together, these components form a robust and explainable pipeline for intrusion detection in IoT environments, one that supports both operational deployment and post-incident forensic analysis.

## 2  Related Work
### 2.1  Explainable Artificial Intelligence (XAI) in Cybersecurity

The integration of explainable AI into cybersecurity models has become increasingly essential as ML systems are deployed in high-stakes, critical environments. XAI aims to enhance trust, auditability, and human understanding of ML outputs, particularly in domains such as threat detection, where false positives and black-box predictions pose substantial operational risks. Arrieta et al. [11] provided a comprehensive taxonomy of XAI approaches, highlighting their potential in domains that require accountability and transparency. In the security-specific domain, Xie et al. [5] reviewed XAI techniques in cybersecurity applications, noting the growing shift from opaque neural models to interpretable methods and the need for model-agnostic interpretability tools. Ghosh et al. [12] also emphasized the need for XAI in threat detection workflows, where security analysts must quickly assess the credibility and origin of alerts. While other explainability methods, such as Grad-CAM [13], have also been applied in deep learning security contexts, and deep learning methods with explainable extensions have also been applied to intrusion detection systems [14], many XAI applications in IDS remain superficial, limited to post-hoc feature ranking without deeper contextualization or class-aware insights.

### 2.2  Distinction from Prior Work

In contrast to prior literature, this work introduces a class-aware explainable AI framework for IoT threat attribution that makes the following key contributions:

- `Multi-class threat modeling with semantic grouping`: Instead of treating each of the 34 original attack types individually or collapsing into a binary setup, we group them into eight high-level, semantically coherent classes: DDoS, DoS, Mirai Botnet, Spoofing, Reconnaissance, Web Exploits & Injection, Brute Force, and Benign Traffic. This structure balances detection accuracy and explainability while allowing consistent labeling for threat attribution.
- `Global and local SHAP-based explainability`: We apply SHAP not only to determine global feature importance but also to visualize per-class beeswarm plots and local instance waterfall plots, uncovering nuanced decision paths that the model follows under different threats.
- `Flow-level behavioral reasoning`: Beyond numerical attribution, we interpret top SHAP-ranked features in the context of network behavior semantics, for example, packet burstiness via IAT, session termination via RST flags, and payload consistency via variance and magnitude. This allows analysts to connect feature changes with operational threat behavior.

## 3  Dataset Overview

While several benchmark data sets have been proposed for intrusion detection, such as UNSW-NB15 [15], these data sets lack large-scale IoT traffic data [16]. The CICIoT2023 dataset is a novel, large-scale, real-time network traffic dataset designed to emulate benign and malicious behavior in Internet of Things (IoT) environments. It captures the complexity and diversity of modern IoT networks, comprising 34 distinct labels that represent various types of cyberattacks, such as DDoS, DoS, Spoofing, Reconnaissance, and malware-based threats like Mirai botnets. The dataset contains over 10 million flows collected across multi-device scenarios, simulating attacks against smart homes and enterprise IoT infrastructure. In this work, a subset of the dataset with about 1 million samples was used. Each sample in the dataset corresponds to a network flow and includes 47 features that encompass packet-level attributes, protocol flags, timing characteristics, and statistical aggregations of flow behavior. The dataset consists of the following major feature groups:

- `Flow Duration & Timing`: Features such as *flow _duration*, *duration*, and *IAT* capture the temporal characteristics of the flow.
- `Header and Protocol Flags`: Binary flags for TCP/IP protocols (e.g., *ACK, SYN, FIN, RST*, etc.), and protocol presence indicators (e.g., *HTTP, DNS, TCP, UDP*)
- `Statistical Aggregates`: Flow-level aggregations such as *Tot sum, AVG, Min, Max, Std, Tot size,* and higher-order metrics such as *Magnitude, Radius, and Variance*.

To improve explainability and reduce semantic overlap during modeling, we reorganized the labels into eight higher-level threat categories based on shared behavioral traits and operational objectives, as seen in Table 1.

Table 1. Mapping of Original ANack Labels to Grouped Threat Categories

| Grouped Class | Original Labels |
| --- | --- |
| Benign Traffic | BenignTraffic |
| DDoS Attacks | DDoS RSTFIN Flood, DDoS ICMP Flood, DDoS SYN Flood, DDoS PSHACK Flood, DDoS TCP Flood, DDoS UDP Flood, DDoS ICMP Fragmentation, DDoS UDP Fragmentation, DDoS ACK Fragmentation, DDoS HTTP Flood, DDoS Slow Loris, DDoS Synonymous IP Flood |
| DoS Attacks | DoS-TCP Flood, DoS-UDP Flood, DoS SYN Flood, DoS HTTP |
| Flood Reconnaissance | Recon Host Discovery, Recon Port Scan, Recon OSScan, Recon Ping Sweep, Vulnerability Scan |
| Spoofing Attacks | DNS Spoofing, MITM |
| ArpSpoofing Brute Force Attacks | Dictionary Brute Force |
| Web Exploits and Injection | XSS, Sql Injection, Command Injection, Uploading Attack, Browser Hijacking, Vulnerability Scan, Backdoor Malware |
| Mirai Botnet Attacks | Mirai Greeth Flood, Mirai Greip Flood, Mirai |
| UDPplain | |

## 4 Methodology

Our approach includes a detailed analysis of flow-level traffic characteristics to uncover patterns, anomalies, and relationships that define normal and malicious behavior. Following this, a data processing pipeline was implemented, which includes feature selection based on insights from analysis, variable transformation (such as encoding for categorical variables and scaling for numerical features), and handling class imbalance. Next, predictive modeling was performed using tree-based classifiers, evaluating their performance through cross-validation and standard classification metrics (accuracy, precision, recall, and F1 score). Finally, SHAP (Shapley Additive Explanations) was applied to explain the model's decisions both at a global and local level. This included the use of class-specific feature attribution (beeswarm plots), instance-level justifications (waterfall plots), and interaction-based reasoning from visual explanations.

### 4.1 Exploratory Analysis of Flow Characteristics

The dataset exhibits a pronounced class imbalance in the eight grouped traffic categories. As shown in the figure below, DDoS attacks overwhelmingly dominate the dataset with over 70% of the total samples, followed by DoS attacks, which represent a sizable but notably smaller share. Other classes, such as Mirai Botnet Attacks, Benign Traffic, and Spoofing Attacks, are significantly underrepresented, while Brute Force, Reconnaissance, and Web Exploits & Injection appear as extreme minority classes. This imbalance poses a modeling challenge, especially for multi-class classification tasks, as models may be biased toward majority classes. To mitigate this issue, we adopt class-weight adjustment in the training process to improve the model's sensitivity to minority attack types without resorting to oversampling or undersampling, which may distort flow-level distributions.

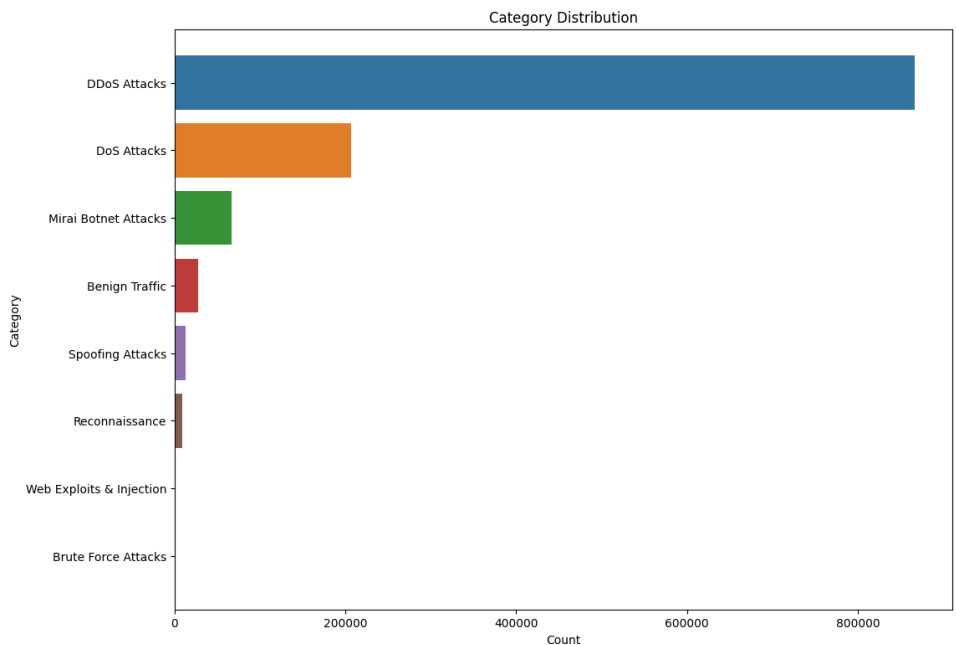

Fig. 1. Class Distribution of Threat Types.

Basic analysis reveals significant skewness and variability in the statistical distribution of flow-based features within the CICIoT2023 dataset. Features like *flow duration*, *Header Length*, and *IAT* exhibit long-tail distributions, suggesting that while most network flows are short-lived and minimal, some carry exceptionally large values, potentially indicative of abnormal activity. Several packet-level and rate-based metrics, such as *Rate, Srate, and Drate*, demonstrate extreme kurtosis, with many flows having near-zero or low values and a few experiencing high spikes. In particular, *Drate* is largely zero in most samples, implying limited downstream traffic in many flows. TCP flag counts (*ack_count , syn_count , rst count , urg count* ) reflect sparse but spiky behavior, where certain flows trigger excessive flags, possibly corresponding to flooding or scan-based intrusions. These features are important candidates for detecting volumetric and protocol-specific attacks. Packet size metrics (*Totsum, Min, Max , AVG, Totsize*) are generally clustered around common control packet lengths (50–60 bytes), but show wide ranges. These deviations could be informative for distinguishing between benign and exploitative communication patterns. Mathematically derived features such as *Magnitude*, *Radius*, *Covariance*, and *Variance* also reflect a high degree of sparsity and occasional outliers. These metrics, especially when large, may signify anomalous payload dynamics or asymmetric flow behaviors. In contrast, features such as *weight* and *Drate*, which show limited variation or constant values, may contribute little to model learning and were considered to be removed.

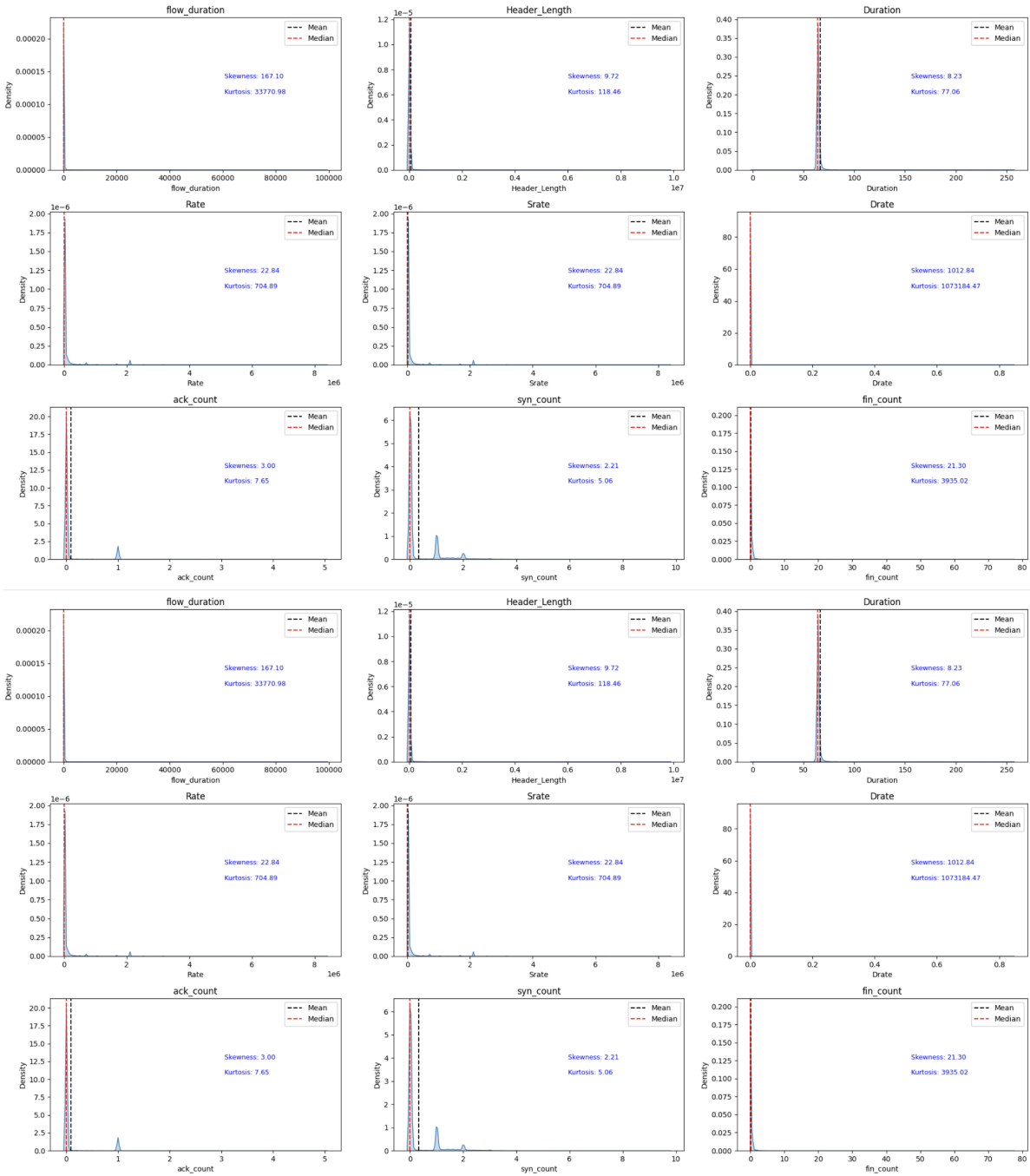

Fig. 2. Class Distribution of Threat Types across different categories.

The correlation matrix revealed several strong relationships between features, notably between groups such as (*Magnitude*, *AVG*, *Max*, *Min*, *Std*, *Totsize*) and (*Radius*, *Variance*), as expected, indicating that they capture overlapping flow characteristics. Furthermore, *IAT* and *Number* showed a correlation with variance, suggesting structural redundancy. Meanwhile, features such as *duration* showed moderate positive correlations with urg_count (r≈0.52) and rst _count (r≈0.75), which may reflect behavioral patterns in flows that are longer or prematurely terminated, offering useful indicators to distinguish certain categories of network attacks. However, many features maintained low pairwise correlations, implying their potential individual utility in classification. To validate these observations, a Variance Inflation Factor (VIF) analysis was performed. The results confirmed serious multicollinearity in variables such as Std (VIF ≈ 16,600), Radius (≈ 16,500), Weight, Number, and IAT, indicating that they may inflate variance in downstream modeling. In particular, *Rate* and *Srate* presented infinite VIFs, which confirms the perfect collinearity. These variables were flagged for removal and transformation in the preprocessing stage to improve model stability and explainability.

The class-wise boxplot analysis of numerical features revealed distinct distributional signatures across different attack categories. Packet-based features such as *RST*, *URG*, and *SYN counts* demonstrated class-specific concentration patterns, particularly for *DDoS* and *DoS*, where most values clustered around zero, indicating minimal flag activity per flow. In contrast, *ACK* and *FIN counts* showed limited variation, which offers little discriminative value. Time-based features, such as *Flow Duration*, *Duration*, and *IAT*, proved especially informative, with DDoS, DoS, and Mirai Botnet flows exhibiting compact, low *IAT* distributions, while other classes had significantly higher interarrival times. Aggregated flow statistics (*Tot _sum*, *Min*, *Max*, *Avg*, *Tot size*) varied widely by class; notably, volumetric attacks such as DDoS and Mirai exhibited compressed distributions with low variance, while more complex or exploratory threats showed a wider spread. Similarly, higher-order statistical metrics (*Magnitude*, *Radius*, *Covariance*, *Variance*, and *Weight*) showed that benign traffic tends to exhibit greater dispersion, especially in variance, while DDoS-like attacks were marked by tighter lower-value ranges. These findings reinforce the hypothesis that different types of attacks have different flow behavior profiles that can be leveraged for effective multiclass threat attribution.

The relationship between categorical features and target threat classes was evaluated using contingency plots and Cramér's V statistic. Key variables such as *Protocol Type*, *ACK _Flag*, *HTTPS*, and *TCP* demonstrated the strongest associations with attack categories, with Cramér's V values exceeding 0.3. These features exhibit distinct frequency distributions between traffic types and may prove valuable in discriminating specific threat classes when considered together with numerical variables. Although *ICMP*, *UDP*, and *SSH* appeared moderately relevant based on statistical measures, visual inspections reveal that these associations may be exaggerated by class imbalance, as their actual distribution is relatively uniform across classes. In contrast, other categorical attributes, including *DHCP*, *ARP*, *SMTP*, and *IRC*, showed weak or negligible relationships with the target variable, both statistically and visually. These features are candidates for removal during feature selection as they contribute little to model performance or explainability.

Collectively, the EDA validates that different classes of threats express themselves through distinctive flow-level and structural behaviors. These patterns reinforce the design of a class-aware interpretability approach by confirming that features carry semantic weight in representing attacker strategies. This alignment between statistical patterns and operational semantics forms the foundation for transparent SHAP-driven threat attribution in the proposed framework.

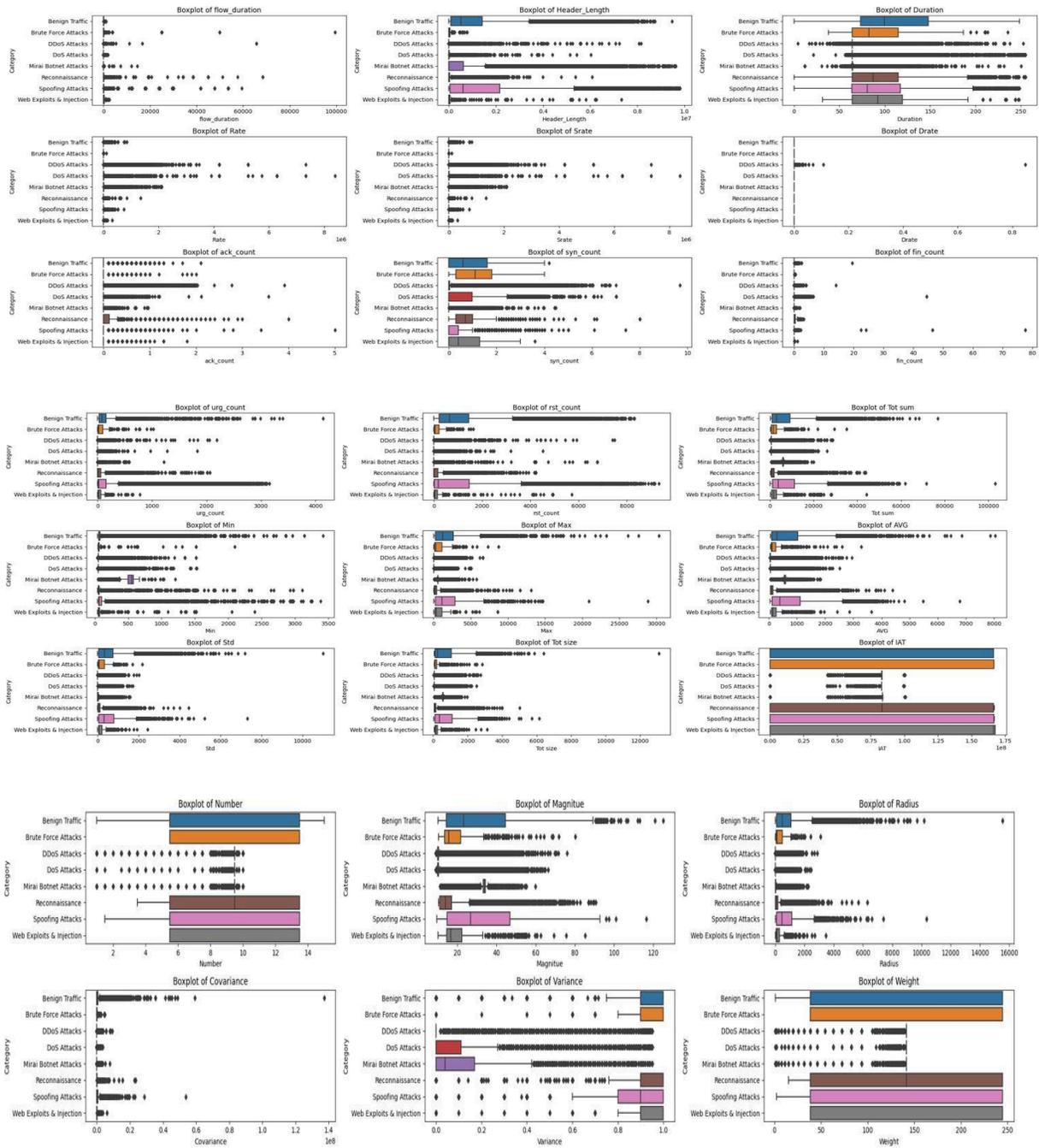

Fig. 3. Distributional Patterns of Threat Types.

## 4.2 Data Processing

To prepare the dataset for effective threat attribution and explainable modeling, we applied several preprocessing steps aimed at reducing noise, optimizing dimensionality, and standardizing input formats. Based on insights from the exploratory analysis, we identified a subset of features that introduced multicollinearity or showed minimal variation between attack classes. Features such as *Srate*, *std*, *number*, *radius*, and *weight* exhibited an extremely high correlation with other variables, suggesting redundancy. Similarly, low-variance features such as *Drate*, various TCP flag counters (*fin flag number*, *syn flag number*, etc.), and rarely used protocol indicators (*DNS*, *SMTP*, *IRC*, etc.) were deemed uninformative for multiclass classification. These were removed to streamline feature space and reduce the risk of model overfitting. Categorical features such as Protocol Type, *TCP*, *UDP*, and others were encoded into a numerical format to ensure compatibility with tree-based learning algorithms. Additionally, the *IAT* feature, originally in microseconds, was scaled down to seconds to improve explanations and reduce its numerical range, preventing it from disproportionately affecting tree splits. The final preprocessed dataset was partitioned into training and validation sets, ensuring that the model evaluation metrics would reflect the performance on unseen samples and preserve generalizability.

## 4.3 Predictive Modeling

To accurately classify and attribute network flows to specific types of IoT threats, we employed a structured predictive modeling pipeline that involved model selection, class imbalance handling, and model refinement. The focus was on interpretable and high-performing tree-based ensemble models. Five widely used ensemble classifiers were considered: Random Forest, XGBoost, LightGBM, Gradient Boosting, and Extra Trees. These models are well-suited for handling high-dimensional tabular data and inherently support feature importance estimation, which aligns with the study's explainability goals. Hybrid multi-layer intrusion detection systems have also been proposed to improve detection precision [17]. We evaluated the models using 4-fold cross-validation with four key metrics: Accuracy, Recall, Precision, and F1-score, computed using a macro-averaging strategy to ensure balanced performance across all threat categories.

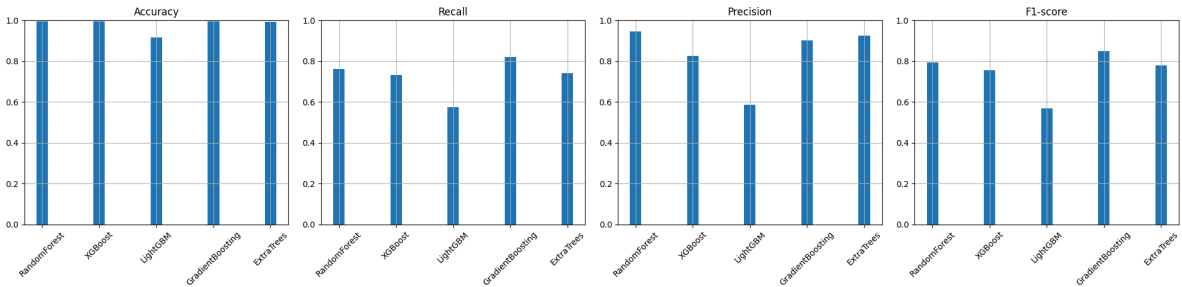

Fig. 4. Model Selection and Evaluation.

The results (as visualized in Figure 4) reveal that while all models perform well in terms of accuracy, they differ significantly in terms of recall and F1-score. Notably, Gradient Boosting achieved the best balance between Recall and F1-score, which are critical for multi-class threat detection, especially under imbalanced conditions. LightGBM underperformed on both precision and recall, while Random Forest and Extra Trees maintained high precision but lower recall. Given its superior balance of predictive power and robustness across metrics, Gradient Boosting was selected as the final model for interpretation. To mitigate bias toward majority classes, we applied a class-weight adjustment strategy during model training. The class weights were computed using a balanced scheme and softly scaled to avoid extreme penalization of frequent classes.

The weights computed were then used to assign sample-specific weights to the training instances, ensuring that the underrepresented classes had greater influence during the loss calculation. This was to enable the final model to generalize more effectively across the eight attack categories, thereby avoiding overfitting to dominant traffic types and improving the model's robustness in detecting minority classes.

```
Classification Report using Adjusted Class Weights
                          precision    recall  f1-score   support

           Benign Traffic       0.90      0.96      0.93      6968
       Brute Force Attacks      0.82      0.56      0.67        91
              DDoS Attacks      1.00      1.00      1.00    216788
               DoS Attacks      1.00      1.00      1.00     51625
       Mirai Botnet Attacks      1.00      1.00      1.00     16798
            Reconnaissance      0.86      0.82      0.84      2267
          Spoofing Attacks      0.90      0.83      0.86      3125
   Web Exploits & Injection      0.72      0.53      0.61       154

                  accuracy                          1.00    297816
                 macro avg      0.90      0.84      0.86    297816
              weighted avg      1.00      1.00      1.00    297816
```

Fig. 5. Classification Report Using Adjusted Class Weight.

Following training with adjusted class weights, the Gradient Boosting model was evaluated on the held-out validation set. The classification report shows improved generalization across all threat classes, particularly for minority categories such as spoofing, reconnaissance, and web exploits, which typically suffer from poor recall in imbalanced settings. The F1-score demonstrated a significant increase compared to earlier baseline models, confirming that class-weighted training improved both precision and recall uniformly across classes. Dominant classes such as DDoS and DoS maintained high performance, while previously underrepresented classes now had notably better coverage. These results affirm the model's capability not only to distinguish between attack types but also to do so fairly.

### 4.4 Explainability with SHAP

We employed SHAP (SHapley Additive exPlanations) to perform both global and local feature attribution. Our approach is inspired by prior SHAP-driven forensic analysis of network flows [18]. This approach provides a model-agnostic explanation by quantifying the contribution of each input feature to the model's output, both overall and per instance. We used the model's native feature importance rankings as a baseline and SHAP summary plots to decompose the model's predictions, showing how each feature contributes to the prediction for each threat class, combining both the magnitude of the feature and the direction of influence. Additionally, we visualized the mean SHAP values across all instances and grouped them by class, providing a clearer understanding of how feature importance varies for different attack types. This helps identify not just the most impactful features overall, but those that are class-discriminatory, vital for multiclass threat attribution.

The global interpretation showed that
- **1. IAT (Inter-Arrival Time) is the most globally influential feature:** Its SHAP values dominated the importance plot, especially in the detection of DoS, DDoS, and reconnaissance attacks. In particular, lower *IAT* values, representing rapid and tightly spaced packet flows, pushed predictions strongly toward malicious classes. This aligns with real-world characteristics of automated attacks, which often exhibit rapid-fire packet transmission compared to the more variable communication seen in benign or user-driven flows.

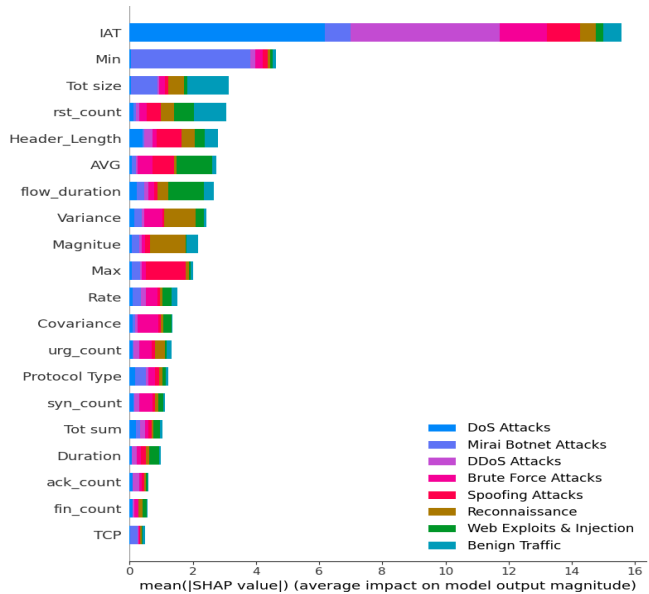

Fig. 6. Average impact on model output magnitude.

- **2. The total size and minimum packet size also contributed significantly to threat classification:**
  These features exhibited large negative SHAP values across several threat types, particularly brute-force, spoofing, and web exploits & Injection. Their importance stems from the tendency of these attacks to generate repetitive, lightweight traffic patterns composed of small packets, such as log-in probes or malicious HTTP requests, making them reliable indicators of suspicious activity.
- **3. Magnitude and Variance captured structural irregularities in packet flow:** These features, which summarize the spread and intensity of packet sizes within flows, were especially impactful for Reconnaissance, Brute Force, and Mirai Botnet detection. High variance was associated with non-benign predictions, reflecting the erratic nature of these attack behaviors. Similarly, low magnitude values, indicative of short and sparse flows, were common in probing activities, as seen in reconnaissance threats.
- **4. Flow Duration and Header Length provided additional temporal and protocol-level signals:**
  Flow Duration showed slightly negative SHAP contributions for attack classes, reinforcing the notion that shorter flows are more indicative of threats such as DDoS and DoS. Interestingly, *Header _Length* had positive contributions to benign predictions, suggesting that standard or well-formed packet headers are more likely to appear in legitimate traffic, while attackers may craft unconventional headers for evasion.
- **5. The RST Count, URG Count, and protocol flags played key roles in class-specific behavior modeling:** *RST* (Reset) packets were notably influential in classes such as spoofing, web injection, and benign, revealing how the disruptions of TCP sessions factor into the classification. *URG count*, which indicates urgent TCP control signals, was particularly relevant in spoofing and injection-based attacks, hinting at manipulated or malformed traffic. These TCP-level cues help the model detect low-level anomalies associated with sophisticated threat strategies.

This global analysis demonstrates that the behavior learned by the model aligns well with theoretical and empirical understandings of network intrusions. It distinguishes threats not just based on feature magnitude, but also based on semantic class-relevant signals such as timing regularity (*IAT*), flow structure (*Tot size*, *Variance*), and protocol behavior (*RST*, *Header Length*). These insights validate the explainability of our SHAP-based approach and provide a strong foundation for explanation and response to class-specific threats. To derive localized, class-aware attribution and visualize how feature values influence individual predictions per attack type, we employed SHAP beeswarm plots. These plots reveal patterns of how low or high values of specific features increase or decrease the likelihood of a class prediction, as well as the variability and directionality of feature influence for each threat category.

The benign traffic class is distinguished by its structural regularity and balanced packet features. High *rst count*, *Header Length*, and metrics such as *AVG*, *Magnitude*, and *Tot size* contribute positively to the prediction of benign flows, indicating a preference for well-formed, voluminous, and consistent communication. In contrast, low *IAT* values reduce the likelihood of benign classification, reflecting the model's sensitivity to bursty or excessive traffic, which is more typical of attacks. For brute-force attacks, the model is guided by high variability and connection-oriented signaling. Features such as variance, *IAT*, *flow duration*, and TCP flags (*urg count*, *syn_count*, *ack flag number* ) dominate, signaling repeated login attempts with irregular packet timing and extended session durations. Negative contributions from Tot size and Tot sum reinforce the idea that these flows are control-heavy but light on actual data. DDoS attacks are identified by their minimalism and uniformity. The model strongly responds to low values in *flow duration*, *Header Length*, *Tot size*, and *Rate*, capturing short-lived, small, and high-frequency traffic. Interestingly, *IAT* has a dual impact; both high and low values affect the predictions, indicating the flexibility of the model in detecting both high-rate and stealthy DDoS patterns. DoS attacks share similarities with DDoS but exhibit more structural consistency. Low *IAT* values are again the most influential, consistent with high-frequency flooding behaviors. The low values in *flow duration*, *Header Length*, and *Tot sum* suggest that these attacks involve repetitive short bursts with limited payloads. The inclusion of covariance and variance with low SHAP values suggests that a lack of traffic variability is also indicative of DoS behavior.

Mirai Botnet attacks are identified through different signatures: large packet sizes, extended inter-arrival times, and steady-flow patterns. Features such as *Min*, *IAT*, and *Tot size* push the model toward the Mirai class. Negative SHAP contributions from *urg count*, *variance*, and the usage of the TCP flag suggest that the model views traditional TCP session markers and high statistical fluctuation as signs of other attack types rather than botnets.

For reconnaissance, the model combines variability with subtle timing and structural cues. The high *variance*, *flow duration*, and the *RST* count help the model recognize scanning activity, while the moderate influence of covariance, *SYN count*, and *Header Length* reflects the controlled but irregular probing style of reconnaissance attempts. The low-magnitude values become more decisive, indicating sensitivity to flow-level subtlety.

Spoofing attacks are identified by longer *IAT*, larger *Header Length*, and lower *Tot size*, *AVG*, and *Max* packet lengths. This indicates that spoofed traffic tends to be minimal, delayed, and possibly crafted with extended headers to simulate legitimacy. *RST* count has a mixed effect depending on the context, reinforcing the idea that failed handshakes or manipulated sessions play a role in spoofing detection. Lastly, Web Exploits & Injections are tied to high *f low _duration*, *Tot _size*, and irregular protocol flag behavior. Features such as *rst count*, *urg count*, and *syn count* become important when viewed with contextual combinations, showing positive or negative effects depending on surrounding traffic patterns. Lower values in Rate, Duration, and *Header Length* often increase prediction confidence, pointing to precise but lightweight attack patterns.

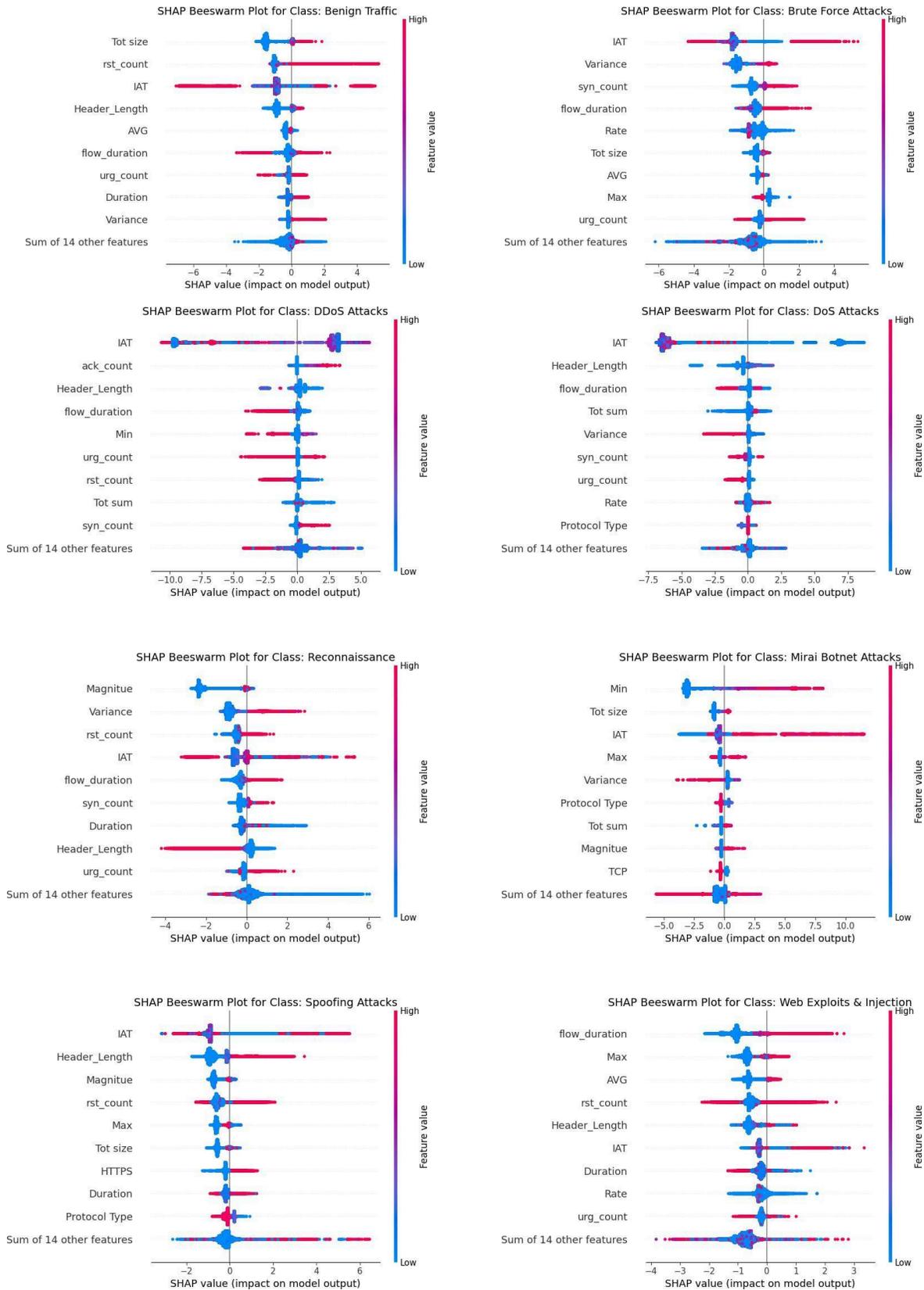

Fig. 7. SHAP Beeswarm plots for Threat Types.

Waterfall plots were also generated for each class using mean feature values. These visualizations illustrate the step-by-step reasoning the model takes to arrive at a specific prediction, showing how the base value changes as each feature contributes positively or negatively to the predicted logit. Together, these explainability tools form the foundation of the class-aware SHAP-based threat attribution framework, enabling both diagnostic insight and trust-building interpretation of model decisions.

To complement the class-specific beeswarm plots, SHAP waterfall plots were used to visualize the average decision trajectory the model follows when classifying each threat type. These plots reveal how the prediction for each class is shifted from the model's expected base value by the cumulative effect of individual features, offering a transparent breakdown of the model's reasoning. For Benign Traffic, the model starts with a base log-odds value of approximately −0.96 and strongly shifts toward a benign classification (f(x) = −5.4) using a collection of features that reflect structured and stable network behavior. Features like *rst count*, *Header Length*, *AVG,* and *Tot size* contribute negatively (toward benign), indicating the model favors legitimate flows that exhibit well-formed sessions, moderate to high packet sizes, and predictable traffic structure. Low burstiness (high IAT) and packet regularity are also key factors. In the case of Brute Force Attacks, the model shifts the prediction sharply from −2.38 to −7.65, confidently identifying the attack. It builds this decision using a combination of temporal irregularity and control flag behavior, notably high IAT, Variance, *urg count*, and TCP signaling (*syn count*, *ack flag_number* ). The lack of positive SHAP values further demonstrates that every input signal reinforces the Brute Force classification, highlighting the model's reliance on session-heavy, low-payload attack patterns. The DDoS Attacks waterfall plot reveals a more moderate shift from E[f(x)] = 5.61 to f(x) = 6.18. This smaller, cumulative nudge results from multiple features such as low *flow duration*, low *Tot size*, and high *Rate*, which together reflect minimal, fast, and repetitive traffic patterns typical of DDoS floods. As seen in the beeswarm plots, a high IAT slightly penalizes classification, revealing that not all DDoS flows exhibit ultra-low inter-packet timing, which demonstrates the model's adaptability to different DDoS attack variants. In DoS Attacks, the model makes a confident non-DoS prediction, dropping from a base of 1.24 to −2.78. IAT drives most of this shift, with high inter-arrival times indicating that the flow does not exhibit the rapid, flood-like behavior typically associated with a DoS attack. The model reinforces this with low values of *flow duration*, *Header Length*, and *Tot sum*, confirming that DoS detection is largely tied to short, repetitive, and time-dense flows.

For Mirai Botnet Attacks, the prediction is confidently negative, shifting from −1.77 to −6.37. The model rules out Mirai by highlighting features like low Min, low IAT, and small Tot size, all of which contrast with the structured, voluminous, and slow-paced traffic typically linked to Mirai botnets in this dataset. This decision path emphasizes the model's capacity to differentiate botnet communication from lightweight or control-plane traffic. The Reconnaissance prediction also shows a sharp decline from −1.28 to −5.36, strongly rejecting this class. The decision is based on a lack of variability and directionality, as low Variance, Magnitude, and Covariance downplay the erratic probing behavior characteristic of reconnaissance scans. The model thus identifies recon flows as highly variable and behaviorally complex, penalizing stable, simple flows.

In Spoofing Attacks, the model similarly declines classification due to short flow durations, low header complexity, and minimal statistical deviation. While certain features like RST count or *f low _duration* might swing predictions under other conditions, their effect here is either neutral or aligned against spoofing. The model consistently requires manipulated timing, header inflation, or connection instability to support a spoofing prediction.

Lastly, for Web Exploits & Injections, the model confidently makes a positive prediction, shifting from −2.11 to −6.56. The decision is driven by long-lived flows, the presence of TCP control signals (RST, URG), and low diversity in packet structure, reflecting the deliberate and session-intensive nature of injection-style threats.

This decision pathway aligns with the SHAP summary and beeswarm analysis, affirming the model's focus on flag-based disruptions and targeted flow manipulation.

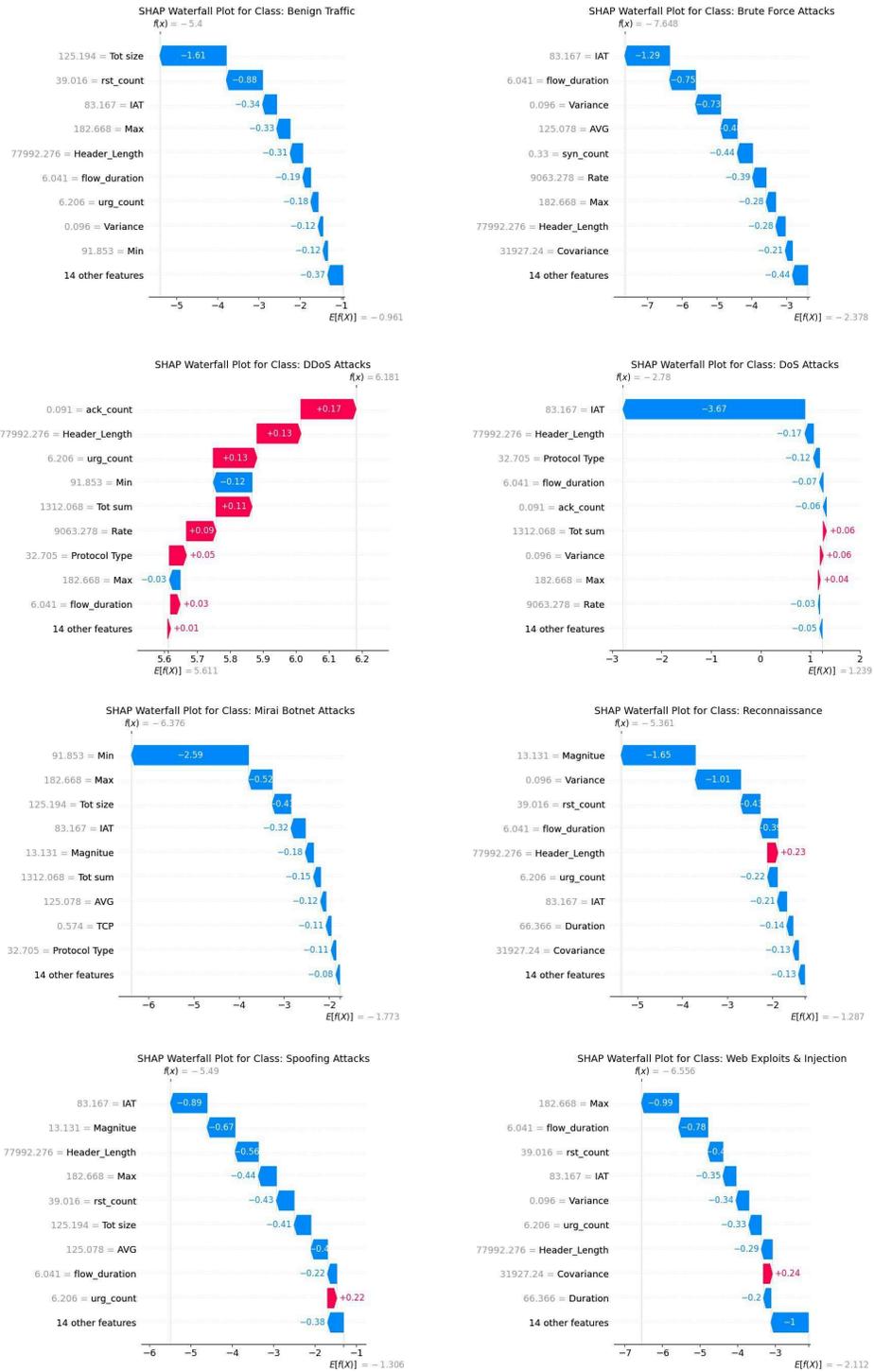

Fig. 8. SHAP Waterfall plots for Threat Types.

# 5 Discussion

## 5.1 What the Model Learns About Threats

The internal logic of the model reveals a structured understanding of different cyber threats in IoT networks. It does not rely on isolated features, but instead infers threat semantics by learning combinations of temporal behavior, packet structure, and flow-level statistics that characterize different attack strategies. A central signal learned by the model is burstiness, measured via the Inter-Arrival Time (IAT). For both DoS and DDoS attacks, low *IAT* values (indicating tightly packed, rapid packet delivery) strongly push the model toward predicting an attack. This aligns with the expected behavior of flooding-based attacks, where the packet frequency is a key weapon. In contrast, in benign flows or Mirai Botnet traffic, higher *IAT* values act as positive indicators, reflecting passive or command-and-control flows with longer delays between transmissions. This shows the model's ability to differentiate between active and stealthy behaviors using timing dynamics. The model also learns to distinguish threats by their packet size characteristics. Small, consistent packet sizes (*Min, Tot size, AVG*) are hallmarks of Brute Force, Web Exploit, and DDoS attacks, which typically do not require payload-heavy communication. In contrast, larger or more variable packet sizes are associated with Mirai Botnet and Benign traffic, where legitimate or structured communications take place. The model interprets packet size not in isolation but in the context of other variables like session duration and flow regularity, revealing an understanding of payload minimalism vs. data richness. Another core dimension is statistical irregularity, captured through features such as *variance*, *magnitude*, and *covariance*. High variability pushes the model toward classes such as Reconnaissance and Brute Force, where inconsistency in flow behavior reflects scanning or trial-and-error attempts. On the other hand, the model penalizes high statistical noise in classes like Benign or Spoofing, where uniform behavior is expected. This shows that the model does not just measure entropy; it contextualizes it based on the class profile. Finally, the model uses protocol-specific flag patterns, such as *RST*, *SYN*, *URG*, and *ACK counts*, to make nuanced distinctions. High *RST counts* push predictions toward Benign or Spoofing, depending on context, reflecting either normal session closures or failed spoof attempts. Meanwhile, multiple *SYN* flags and long *flow duration* help the model recognize Brute Force traffic, which involves repeated connection attempts. This shows the model's grasp of low-level protocol behaviors and how they map to attack mechanics. In general, the model's decisions reflect a multidimensional interpretation of flow-level telemetry. It internalizes the semantics of timing, volume, and protocol behavior and tailors its predictions to combinations of features that align with known threat archetypes. These insights not only validate the predictive robustness of the model but also provide explainable threat characteristics for security analysts.

## 5.2 **Limitations**

Although the model demonstrates a strong ability to characterize and distinguish between IoT threats based on flow behavior, some important limitations remain that affect its reliability and deployment in real-world environments. First, the model shows sensitivity to highly class-specific patterns, which, while beneficial for explainability, may result in overfitting to known threat behaviors. As revealed by the SHAP analysis, the model's decisions are often shaped by narrowly defined thresholds, such as specific values for inter-arrival time, packet variance, or header length. Although these learned thresholds are useful for attribution within the training distribution, they may not generalize well to novel or evolving attacks that deviate from historical traffic profiles. For example, slight changes in the delivery rate or structure of a new DDoS variant could cause the model to misclassify the traffic as benign or associate it with another class entirely. Furthermore, the dataset used (CICIoT2023) is a static and prelabeled dataset, which limits the model's exposure to adaptive or multistage attacks often seen in real-world IoT networks. The absence of continuous feedback, temporal context, or adversarial dynamics means that while the model captures the semantics of the current flow level well, its generalization to unseen or adaptive threat vectors remains uncertain. In addition, many of the features it relies on, such as cumulative variance, packet flow duration, or aggregate flag counts, require observing a full session or a substantial chunk of network activity. This makes the current implementation less suitable for proactive detection in production environments. To transition this work into a deployable IDS (Intrusion Detection System), further optimization for streaming or incremental learning would be needed. Despite these limitations, the study offers a compelling step toward class-aware and explainable threat detection in IoT security, especially for environments where transparency, trust, and forensic traceability are essential.

### 5.3 Suggestions for Improvement

Based on the insights and limitations outlined, several improvements can be made to enhance both the robustness and the applicability of the proposed threat attribution framework. One promising direction is the integration of semi-supervised learning to reduce dependence on fully labeled data. Given that real-world IoT traffic is vast and often lacks precise annotation, leveraging unlabeled flows through methods such as self-training, contrastive learning, or autoencoder-based pretraining could help generalize the model to unseen or evolving attack patterns. This would also reduce overfitting to predefined threat signatures and improve performance in settings where threats evolve faster than labeling pipelines. Another vital improvement is the incorporation of online learning or streaming-aware algorithms to support real-time threat attribution. While the current setup operates in batch mode, adapting the model to process live flows incrementally (e.g., using time-windowed aggregation or sketch-based approximations) would allow for deployment in real network environments with low latency and memory constraints. Real-time anomaly detection frameworks, such as river-based models or continual learning architectures, could be investigated for this purpose. In parallel, embedding real-time explainable AI (XAI) capabilities into the inference pipeline can significantly enhance operational transparency. Tools like SHAP KernelExplainer or LIME can be optimized for streaming scenarios by focusing on top-k features or using adaptive sampling to reduce computational overhead. This would allow security analysts to not only receive predictions but also understand why an alert was triggered, instantly, and at scale. Finally, moving beyond flow-level data, integrating cross-modal signals (e.g., device identity, network topology, historical behavior), and deploying graph-based models for joint reasoning could significantly improve performance in heterogeneous IoT environments, where device interactions often exhibit structured dependencies. Together, these enhancements could transform the current framework into a comprehensive, adaptive, and real-time IoT threat intelligence system, scalable, explainable, and ready for operational use in critical infrastructure security.

## 6 Conclusion

This work aimed to build an interpretable, class-aware framework for IoT threat attribution using flow-based telemetry and SHAP-based explainability. Using the CICIoT2023 dataset, we addressed the complex challenge of multiclass intrusion detection in heterogeneous large-scale IoT environments by modeling grouped threat types and analyzing their flow behavior semantics. Through exploratory data analysis, model optimization, and a thorough application of SHAP at both global and local levels, we were able to uncover the nuanced decision-making patterns that drive predictive classification in eight major threat categories. Our findings show that machine learning models can internalize fine-grained traffic characteristics such as packet burstiness, flow duration, structural variance, and TCP flag behavior to distinguish between different types of IoT attacks. The SHAP-based analysis provided both global insights (e.g., the dominance of Inter-Arrival Time and packet size in driving decisions) and local, class-specific reasoning (e.g., how Brute Force attacks are recognized by flow repetition and flag activity, while Mirai Botnet attacks are driven by large, structured payloads). These insights not only validate the learning capacity of the model but also bring transparency to how it reasons about various threat behaviors in a real-world IoT context. Importantly, this work contributes to the growing need for trustworthy AI in cybersecurity. By enabling visibility into what features drive model predictions, we support a shift from black-box intrusion detection to systems that are auditable, interpretable, and aligned with operational reasoning. This transparency is crucial in sensitive environments such as smart homes, healthcare, and industrial control systems, where misclassifications carry high consequences. Looking ahead, this work provides a basis for further advancements in explainable security models, particularly toward real-time deployment, adaptive learning, and expert-in-the-loop validation. Ultimately, by combining accurate threat detection with semantic explainability, this research moves us closer to building AI systems that not only perform well but also earn the trust of analysts, engineers, and society at large.